\documentclass[letter,twocolumn]{jpsj3}

\usepackage{txfonts}
\usepackage{amsmath}
\usepackage{amssymb}
\usepackage{graphicx}

\title{Exponential Speedup of Quantum Annealing by Inhomogeneous Driving of the Transverse Field}

\author{Yuki Susa\thanks{susa@stat.phys.titech.ac.jp}, Yu Yamashiro\thanks{yamashiro@stat.phys.titech.ac.jp}, Masayuki Yamamoto\thanks{yamamoto@stat.phys.titech.ac.jp}, and Hidetoshi Nishimori}
\inst{Department of Physics, Tokyo Institute of Technology, Meguro-ku, Tokyo, 152-8551, Japan}

\abst{We show, for quantum annealing, that a certain type of inhomogeneous driving of the transverse field erases first-order quantum phase transitions in the $p$-body interacting mean-field-type model with and without longitudinal random field. Since a first-order phase transition poses a serious difficulty for quantum annealing (adiabatic quantum computing) due to the exponentially small energy gap, the removal of first-order transitions means an exponential speedup of the annealing process. The present method may serve as a simple protocol for the performance enhancement of quantum annealing, complementary to non-stoquastic Hamiltonians.}

\begin{document}
\maketitle

Quantum annealing (QA) is a quantum-mechanical metaheuristic for combinatorial optimization problems \cite{kadowaki1998quantum,Brooke1999,Farhi2001,Santoro2002,Santoro2006,Das2008,Tanaka_book2017,AlbashLidar2017}. Recent studies show that QA may also be useful for sampling from a Boltzmann-like distribution of the Ising model, and this could lead to a novel approach to machine learning tasks \cite{Biamonte2017}.

One of the theoretical bottlenecks of QA is the existence of a quantum phase transition in the course of its time development. If the system closely follows the instantaneous ground state, the time necessary for such an adiabatic process is proportional to a polynomial of the inverse of the energy gap according to the adiabatic theorem of quantum mechanics \cite{albash2016adiabatic}. If a first-order phase transition exists at some point of time evolution, the energy gap closes exponentially as a function of the system size, and consequently the computation time increases exponentially. This means that the problem is difficult to solve. A second-order quantum phase transition is accompanied by a polynomially-closing gap, which does not pose a threat to QA.
It is likely that most practically interesting problems of combinatorial optimization have a first-order phase transition as long as they are formulated in terms of the conventional transverse-field Ising model suitable for QA. It is therefore imperative to find ways to mitigate this serious problem of first-order phase transitions.

There have been a number of attempts in this direction, among which the use of non-stoquastic Hamiltonians is a prominent example \cite{farhi2002quantum,crosson2014different,nishimori2017exponential,hormozi2017nonstoquastic} although its hardware realization is non-trivial. In the present paper, we analyze another simpler approach of inhomogeneous driving of the transverse field, in which the strength of the transverse field is turned off sequentially from one spin to the next.

Rams {\it et al.} showed, partly analytically and mostly numerically, for the one-dimensional transverse-field Ising model with random interactions that an inhomogeneous field driving yields better results for the residual energy than the conventional uniform driving does \cite{rams2016inhomogeneous}. We show analytically for mean-field-type models that inhomogeneous driving can completely remove quantum phase transitions and thus exponentially accelerate QA. This result holds even for those models in which the method of non-stoquastic Hamiltonians does not work.

The first problem that we discuss is the simple $p$-spin model \cite{jorg2010energy,seki2012quantum,bapst2012quantum,seoane2012many,susa2017relation},
\begin{equation}
    \hat{H}_0=-N \left( \frac{1}{N}\sum_{i=1}^N \hat{\sigma}_i^z \right)^p, \label{H0}
\end{equation}
where $p$ is an integer greater than or equal to 2. The number of spins (sites or qubits) is denoted as $N$, and $\hat{\sigma}_i^z$ is the $z$ component of the Pauli operator for the $i$th spin.  

In traditional QA, we add a transverse field to the above classical Ising Hamiltonian, which is identified as the cost function in the context of combinatorial optimization. Then the total Hamiltonian reads
\begin{equation}
    \hat{H}(s)=s\hat{H}_0 -(1-s)\sum_{i=1}^N \hat{\sigma}_i^x, 
    \label{Hs}
\end{equation}
where $\hat{\sigma}_i^x$ is the $x$ component of the Pauli operator and $s$ is a time-dependent parameter to control the evolution of the system. The adiabatic process of QA starts from $s=0$ at time $t=0$ and ends with $s=1$ at time $t=t_0$. It is known that the present system (\ref{Hs}) undergoes a first-order quantum phase transition at zero temperature if $p\ge 3$ \cite{jorg2010energy}. The minimum energy gap $\Delta$ of Eq. (\ref{Hs}) between the ground state and the first excited state decreases exponentially as a function of the system size $N$ at the transition point. Consequently, the computation time $t_0$ for adiabatic evolution grows exponentially according to the adiabatic condition of quantum mechanics \cite{albash2016adiabatic},
\begin{equation}
    t_0 \gg \frac{\left|\langle 1 |\displaystyle\frac{d\hat{H}}{dt}|0\rangle \right|}{\Delta^2},
    \label{adiabatic}
\end{equation}
where $|0\rangle$ and $|1\rangle$ are the instantaneous ground state and the first excited state, respectively. It is therefore necessary to find ways to reduce the rate of gap closing from an exponential function, $\Delta \propto e^{-aN}~(a>0)$, which is characteristic of first-order transitions, to a polynomial $\Delta \propto N^{-b}~(b>0)$ for second-order transitions, or even better, to a constant.

With this goal in mind, we modify the total Hamiltonian to
\begin{equation}
    \hat{H}(s, \tau)=s \hat{H}_0-\sum_{i=1}^{N(1-\tau)}\hat{\sigma}_i^x,
    \label{Hst}
\end{equation}
where $\tau$ is an additional time-dependent parameter satisfying $0\le \tau \le 1$. In the traditional case of Eq. (\ref{Hs}), the transverse field is applied to all spins uniformly, and its amplitude $1-s$ is decreased as a function of time as $s$ increases from 0 to 1. In the new formulation (\ref{Hst}), both $s$ and $\tau$ are controlled as functions of time, starting from $s=\tau=0$ at $t=0$ and ending with $s=\tau=1$ at $t=t_0$. The initial Hamiltonian, therefore, has just the transverse field, and the final Hamiltonian is the Ising model $\hat{H}_0$, both in agreement with the traditional protocol.

Equation (\ref{Hst}) indicates that the transverse field is applied only to $N(1-\tau)$ spins, and this number decreases as time proceeds with $\tau$ increasing towards 1. In other words, the transverse field is turned off one by one, starting from spin $i=N$ and ending with spin $i=1$ at $\tau=1$. In this way, the transverse field is driven inhomogeneously.

Strictly speaking, the parameter $\tau$ can take only discrete values for finite $N$, since the upper limit of the summation in Eq. (\ref{Hst}), $N(1-\tau)$, should be an integer. In addition, if we suddenly turn off the transverse field applied to a site, as implied in Eq. (\ref{Hst}) with discrete values of $\tau$, the derivative of the Hamiltonian in the numerator of the adiabatic condition (\ref{adiabatic}) diverges. This problem can be circumvented by using a continuous, piecewise-differentiable function to represent the strength of the transverse field for each spin. We defer the discussion on this point to the last part of the paper and proceed to the presentation of our results for the Hamiltonian (\ref{Hst}).

\begin{figure}[t]
\centering
\includegraphics[width=0.7\linewidth]{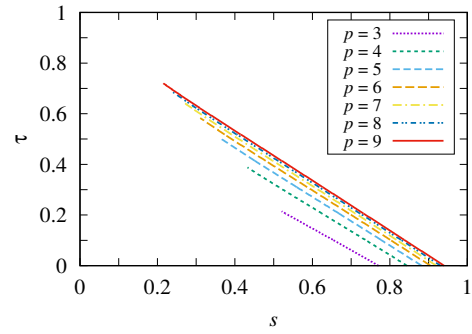}
\caption{(Color online) Ground-state phase diagram. Each line represents a series of first-order phase transitions for a given value of $p$.}
\label{fig:phase_diagram}
\end{figure}

It is straightforward to solve the equilibrium statistical mechanics of the Hamiltonian (\ref{Hst}) by the standard method of Trotter decomposition and the static approximation \cite{jorg2010energy,seki2012quantum,bapst2012quantum,seoane2012many}.  The resulting free energy at finite temperature is
\begin{align}
&f(m; s,\tau) \nonumber\\
&~=(1-\tau)\left\{ (p-1)sm^p -T\log 2\cosh \beta\sqrt{(spm^{p-1})^2+1}\right\}
\nonumber\\
&\hspace{1cm}+\tau\left\{(p-1)sm^p -T\log 2\cosh (\beta spm^{p-1})\right\},
\label{f_Tfinite}
\end{align}
where $m$ is the order parameter for magnetization along the $z$ axis, and $\beta$ is the inverse temperature.  The zero-temperature limit of this free energy is
\begin{align}
 f_0(m; s,\tau)
 &~=(1-\tau)\left\{ (p-1)sm^p-\sqrt{(spm^{p-1})^2+1}\right\} \nonumber \\
 &\hspace{1cm}+\tau\left\{(p-1)sm^p -sp m^{p-1}\right\},
\label{f_T0}
\end{align}
where we assume $m\ge 0$.
Minimization of this zero-temperature free energy with respect to $m$ leads to the phase diagram depicted in Fig. \ref{fig:phase_diagram}.
For a given value of $p$, there exists a line of first-order phase transitions extending from a point on the axis $\tau=0$ towards the middle of the phase diagram. It is remarkable that all these lines terminate before they reach one of the axes, $s=0$ or $\tau=1$. Thus, there exists a path which starts at $s=\tau=0$ and ends at $s=\tau=1$ without encountering a phase transition. This means that we can avoid phase transitions altogether, both first order and second order, and consequently the energy gap stays constant even in the limit of large $N$.

The location of the critical point $s_{\rm c}, \tau_{\rm c}$, where the line of the first-order transitions terminates for a given value of $p$, can be identified analytically by using the standard Landau theory of phase transitions, i.e., by the condition that the coefficients of the expansion of the free energy (\ref{f_T0}) around its minimum at $m=m_{\rm c}$ vanish to third order \cite{Nishimori_book}. The result is
\begin{equation}
    \tau_{\rm c}=\frac{1}{1+\displaystyle\sqrt{\frac{27(p-1)}{4(p-2)^3}}},\quad
    s_{\rm c}=\frac{1}{pm_{\rm c}^{p-1}\sqrt{1-{m_{1\rm c}}^2}/m_{1\rm c}},
\end{equation}
where $m_{1\rm c}=\sqrt{(p-2)/(3(p-1))}$ and $m_{\rm c}=\tau_{\rm c}+(1-\tau_{\rm c})m_{1\rm c}$. This equation shows that the critical point is located in the middle of the phase diagram for any finite $p\ (\ge 3)$, which means that there exists a path connecting the starting and ending points, $s=\tau=0$ and $s=\tau=1$, respectively, without crossing a phase transition.

\begin{figure}[t]
\centering
\includegraphics[width=\linewidth]{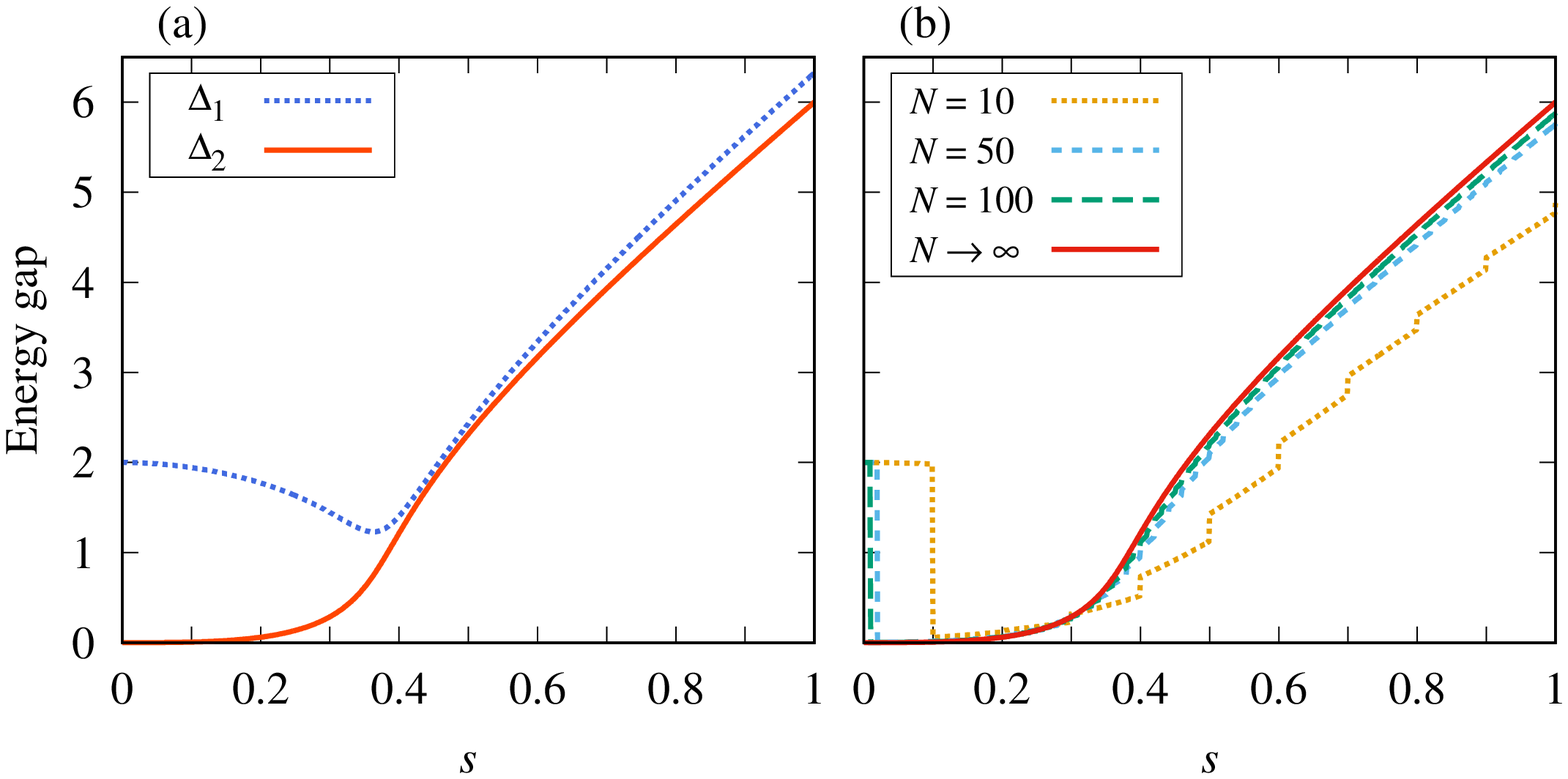}
\caption{(Color online) Energy gap as a function of $s (=\tau)$ for $p=3$ by semiclassical analysis (a) and numerical diagonalization (b). The latter includes the result for the first excited state by the semiclassical analysis in (a).}
\label{fig:gap}
\end{figure}

To reinforce this conclusion, we calculated the energy gap both analytically and numerically. Since the system is of mean-field-type and is therefore semi-classical, it is straightforward to apply the well-established technique to evaluate quantum fluctuations around the classical state \cite{filippone2011quantum,seoane2012many}. To be explicit, we adopt the parameterization of a path $\tau=s^r$ connecting $s=\tau=0$ and $s=\tau=1$ with a parameter $r$ to control the shape of the path. We rewrite the Hamiltonian (\ref{Hst}) in terms of two giant spin operators,
\begin{equation}
    \hat{S}_1^{z,x}=\frac{1}{2}\sum_{i=1}^{N(1-s^r)} \hat{\sigma}_i^{z,x},\quad \hat{S}_2^{z,x}=\frac{1}{2}\sum_{i=N(1-s^r)+1}^{N} \hat{\sigma}_i^{z,x}
\end{equation}
as
\begin{equation}
    \hat{H}(s,\tau)=-sN\left\{\frac{2}{N}\big(\hat{S}_1^z +\hat{S}_2^z\big)\right\}^p
    -2 \hat{S}_1^x.
\end{equation}
We regard these operators as classical vectors in the zeroth approximation for sufficiently large $N$, and subsequently take into account the quantum fluctuations around the classically stable directions via an expansion of the Holstein-Primakoff transformation to the quadratic order in boson operators, precisely as is done traditionally \cite{filippone2011quantum,seoane2012many}. The result is
\begin{align}
\hat{H}(s,\tau =s^r)\approx Ne+\gamma+\frac{\delta}{2}\left(\sqrt{1-\epsilon^2}-1\right)+\Delta_1 \hat{b}_1^{\dagger}\hat{b}_1+\Delta_2\hat{a}_2^{\dagger}\hat{a}_2,
\end{align}
where $e$ is the classical ground-state energy per spin and $\Delta_1$ and $\Delta_2$ represent the quantum fluctuations. These and other symbols in the above equation are defined in terms of the original parameters as
\begin{subequations}
\begin{align}
e&:=-s\{s^r+(1-s^r)\cos \theta_0\}^p-(1-s^r)\sin \theta_0, \\
\delta &:=\Delta_2 \cos\theta_0+2\sin \theta_0+2\gamma, \\
\gamma&:=-\frac{1}{2}sp(p-1)\{s^r+(1-s^r) \cos\theta_0\}^{p-2}(1-s^r)\sin^2\theta_0,\\
\Delta_2&:=2sp \{s^r+(1-s^r)\cos\theta_0\}^{p-1},\\
\epsilon&:=-\frac{2\gamma}{\delta},\\
\Delta_1&:=\delta \sqrt{1-\epsilon^2},
\end{align}
\end{subequations}
where the angle $\theta_0$ is chosen to minimize the energy $e$.
The energy gap between the ground state and the first excited state is the smaller among $\Delta_1$ and $\Delta_2$. In Fig. \ref{fig:gap}, we show $\Delta_1$ and $\Delta_2$ as functions of $s$ under the parameterization $\tau=s^r$ for $p=3$ and $r=1$, which is a path that avoids the line of the first-order phase transitions in the phase diagram (Fig. \ref{fig:phase_diagram}).
In the present case, $\Delta_2$ is the lowest energy gap, which is in very good agreement with the results of the direct numerical diagonalization for finite systems with $N$ up to 100. It should be noted that the above semi-classical analysis gives the energy gap in the large-$N$ limit. We therefore safely conclude that the phase diagram in Fig. \ref{fig:phase_diagram} properly describes the system behavior in the limit of large system size.

Essentially the same result is obtained for the random-field Ising model
\begin{align}
\label{eq:RFIM}
\hat{H}_0=-N\left(\frac{1}{N} \sum_{i=1}^{N} \hat{\sigma}_i^z\right)^p-\sum_{i=1}^N h_i \hat{\sigma}^z_i,
\end{align}
where the longitudinal $h_i$ is applied to each site $i$, whose distribution is either binary $h_i=\pm h_0$ or Gaussian, both with a vanishing mean. It is known that this is a difficult problem for QA in the sense that even the introduction of an $XX$ interaction, which makes the Hamiltonian non-stoquastic, fails to reduce first-order transitions to second order, in contrast to the simple $p$-spin model without randomness, Eq. (\ref{Hs}) \cite{ichikawa2014}.

\begin{figure}[t]
\centering
\includegraphics[width=0.55\linewidth]{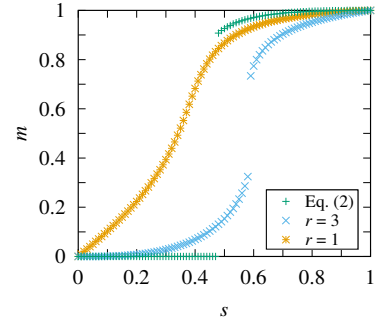}
\caption{(Color online) Magnetization as a function of $s$ under the parameterization $\tau =s^r$ as well as the case of Eq. (\ref{Hs}) for the random-field Ising model with $p$-body interactions with $p=3$ and the strength of the random field $h_0=0.5$. The first order transition that exists when Eq. (\ref{Hs}) and $r=3$ disappears for $r=1$.}
\label{fig:random_field}
\end{figure}

We analyzed this problem by the same method as above, and a part of the results is displayed in Fig. \ref{fig:random_field}. It clearly shows that the first-order transition in the case of a uniform field (Eq. (\ref{Hs})) and an inhomogeneous field $\tau=s^r$ with $r=3$ disappears when $r=1$.
We observed similar phenomena for various combinations of the parameters, $p, r$, and $h_0$ as well as the variance of the Gaussian distribution. Since the entire set of data is extensive, we defer presentation of them to a subsequent paper and just state here that the inhomogeneous driving of the transverse field succeeds in removing the first-order phase transitions for a wide range of parameters in the random-field Ising model.

Let us now turn our attention to the way the transverse field is driven as a function of time. In Eq. (\ref{Hst}), the transverse field is turned off one by one starting from site $N$ ending at site 1, discretely under the assumption that $\tau$ takes discrete values such that the upper bound of the summation $N(1-\tau)$ is an integer. This makes the theoretical analysis easier but is not necessarily very desirable from the viewpoint of the adiabatic condition. In Eq. (\ref{adiabatic}), a sudden change of the Hamiltonian leads to a divergent numerator. This problem can be circumvented as follows.

\begin{figure}[t]
\centering
\includegraphics[width=0.7\linewidth]{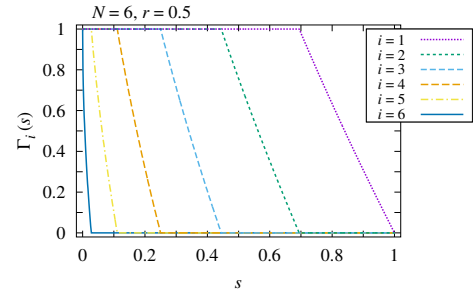}
\caption{(Color online) Continuous function $\Gamma_i(s)$ to interpolate a finite transverse field and a vanishing field at each $i$.}
\label{fig:continuous_function}
\end{figure}

Under the parameterization $\tau=s^r$, we use the following form of the coefficient of the transverse field, instead of the second term of Eq. (\ref{Hst}),
\begin{subequations}
\label{eq:sequential_driver}
\begin{align}
\hat{V}(s)&=-\sum_{i=1}^N\Gamma_i(s)\hat{\sigma}_i^x, \\
\Gamma_i(s) &=
\left\{
\begin{array}{ccl}
1 & \textrm{if} & s<s_i, \\
N(1-s^r)+(1-i) & \textrm{if} & s_i\leq s \leq s_{i-1}, \\
0 & \textrm{if} & s_{i-1}<s,
\end{array}
\right.
\end{align}
\end{subequations}
where  $s_i=(1-i/N)^{1/r}$. As depicted in Fig. \ref{fig:continuous_function}, $\Gamma_i(s)$ is a continuous, piecewise differentiable function. Its maximum slope is proportional to $N$.
In the limit $N\to\infty$, the present inhomogeneous driving reduces to the previous case of Eq. (\ref{Hst}), since the separation between $s_{i-1}$ and $s_i$ shrinks to zero. The time derivative of the Hamiltonian is proportional to $N^2$, and the computation time for an adiabatic process is proportional to $N^2$ with a finite value of energy gap for an appropriately chosen value of $r$.

\begin{figure}[t]
\centering
\includegraphics[width=\linewidth]{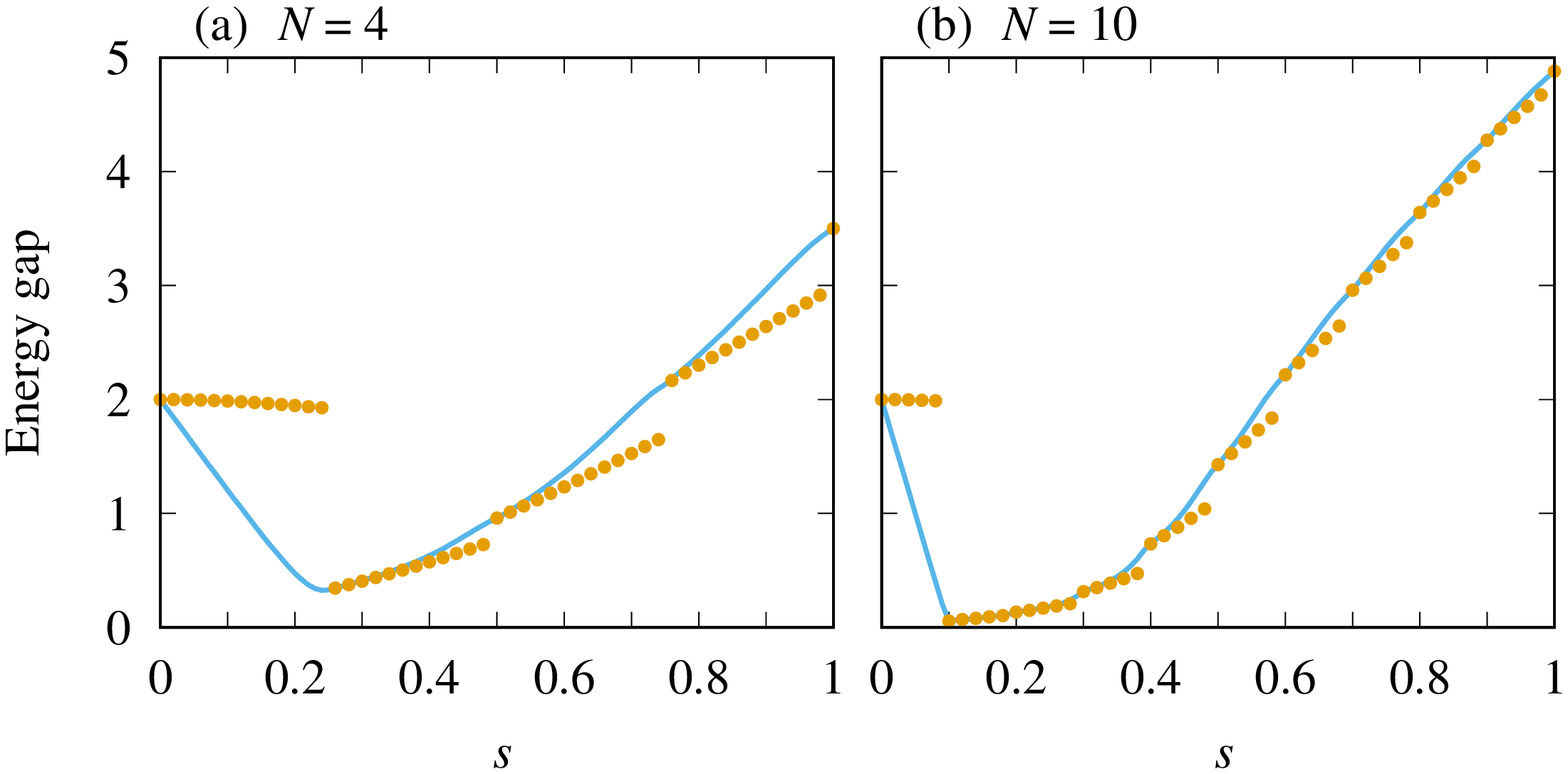}
\caption{(Color online) Energy gap for two types of driving of the transverse field with $p=3$ and $\tau=s$. The dots are for the discrete driving of Eq. (\ref{Hst}) and the continuous curves are for Eq. (\ref{eq:sequential_driver}).}
\label{fig:driving_spectrum}
\end{figure}

We have confirmed the above reasoning by comparing the energy spectra for the two types of inhomogeneous driving, Eqs. (\ref{Hst}) and (\ref{eq:sequential_driver}). The result for the energy gap is shown in Fig. \ref{fig:driving_spectrum}, where the dots denote the energy eigenvalues for Eq. (\ref{Hst}) and the continuous curves are for Eq. (\ref{eq:sequential_driver}).
It is very plausible that the two cases coincide when $N$ is sufficiently large. We are therefore confident that our analysis usig Eq. (\ref{Hst}) is legitimate.  

We have shown that inhomogeneous driving of the transverse field can remove first-order phase transitions of the transverse-field Ising models which exist in the case of a uniform transverse field. This leads to an exponential speedup of QA, because the energy gap remains finite even in the limit of large system size. We have analyzed the simple case of the uniformly interacting $p$-spin model as well as the random-field Ising model. It is not easy to understand intuitively why such a simple protocol is effective in removing phase transitions. One of the possibilities may be that a phase transition is a cooperative phenomenon involving all degrees of freedom, and inhomogeneity of the field would jeopardize the cooperation between different parts of the system. If this crude picture captures some of the essential features of the present scheme, a similar phenomenon might be observed in more complex systems, such as spin glasses with finite connectivity (e.g. in finite spatial dimensions). It would be worth the effort to study many other cases. The present scheme is also attractive from an experimental perspective, since the implementation should be easier than that of non-stoquastic Hamiltonians involving, e.g., $XX$ interactions.

This work was partially funded by the ImPACT Program of the Council for Science, Technology and Innovation, Cabinet Office, Government of Japan, and by the JSPS KAKENHI Grant No. 26287086. The research is based upon work partially supported by the Office of the Director of National Intelligence (ODNI), Intelligence Advanced Research Projects Activity (IARPA), via the U.S. Army Research Office contract W911NF-17-C-0050. The views and conclusions contained herein are those of the authors and should not be interpreted as necessarily representing the official policies or endorsements, either expressed or implied, of the ODNI, IARPA, or the U.S. Government. The U.S. Government is authorized to reproduce and distribute reprints for Governmental purposes notwithstanding any copyright annotation thereon. For numerical calculations, we used the QuTIP library. \cite{JOHANSSON20121760,JOHANSSON20131234} 


\begin{thebibliography}{10}

\bibitem{kadowaki1998quantum}
T. Kadowaki and H. Nishimori, Phys. Rev. E \textbf{58}, 5355 (1998).

\bibitem{Brooke1999}
J. Brooke, D. Bitko, T. F. Rosenbaum, and G. Aeppli, Science \textbf{284}, 779 (1999).

\bibitem{Farhi2001}
E. Farhi, J. Goldstone, S. Gutmann, J. Lapan, A. Lundgren, and D. Preda, Science \textbf{292}, 472 (2001).

\bibitem{Santoro2002}
G. E. Santoro, R. Marton\'{a}k, E. Tosatti, and R. Car, Science \textbf{295}, 2427 (2002).

\bibitem{Santoro2006}
G. E. Santoro and E. Tosatti, J. Phys. A \textbf{39}, R393 (2006).

\bibitem{Das2008}
A. Das and B. Chakrabarti, Rev. Mod. Phys. \textbf{80}, 1061 (2008).

\bibitem{Tanaka_book2017}
S. Tanaka, R. Tamura, and B. K. Chakrabarti, {\em {Quantum Spin Glasses,
  Annealing, and Computation}} (Cambridge University Press, 2017).

\bibitem{AlbashLidar2017}
T. Albash and D. A. Lidar, arXiv:1705.07452.

\bibitem{Biamonte2017}
J. Biamonte, P. Wittek, N. Pancotti, P. Rebentrost, N. Wiebe, and S. Lloyd, Nature \textbf{549}, 195 (2017).

\bibitem{albash2016adiabatic}
T. Albash and D. A. Lidar, arXiv:1611.04471.

\bibitem{farhi2002quantum}
E. Farhi, J. Goldstone, and S. Gutmann, arXiv:quant-ph/0208135.

\bibitem{crosson2014different}
E. Crosson, E. Farhi, C. Y.-Y. Lin, H.-H. Lin, and P. Shor, arXiv:1401.7320.

\bibitem{nishimori2017exponential}
H. Nishimori and K. Takada, Front. ICT \textbf{4}, 2 (2017).

\bibitem{hormozi2017nonstoquastic}
L. Hormozi, E. W. Brown, G. Carleo, and M. Troyer, Phys. Rev. B \textbf{95}, 184416 (2017).

\bibitem{rams2016inhomogeneous}
M. M. Rams, M. Mohseni, and A. del Campo, New J. Phys. \textbf{18}, 123034 (2016).

\bibitem{jorg2010energy}
T. J{\"o}rg, F. Krzakala, J. Kurchan, A. C. Maggs, and J. Pujos, Europhys. Lett. \textbf{89}, 40004 (2010).

\bibitem{seki2012quantum}
Y. Seki and H. Nishimori, Phys. Rev. E \textbf{85}, 051112 (2012).

\bibitem{bapst2012quantum}
V. Bapst and G. Semerjian, J. Stat. Mech. \textbf{2012}, P06007 (2012).

\bibitem{seoane2012many}
B. Seoane and H. Nishimori, J. Phys. A \textbf{45}, 435301 (2012).

\bibitem{susa2017relation}
Y. Susa, J. F. Jadebeck, and H. Nishimori, Phys. Rev. A \textbf{95}, 042321 (2017).

\bibitem{Nishimori_book}
H. Nishimori and G. Ortiz, {\em Elements of Phase Transitions and Critical
  Phenomena} (Oxford University Press, 2011).

\bibitem{filippone2011quantum}
M. Filippone, S. Dusuel, and J. Vidal, Phys. Rev. A \textbf{83}, 022327 (2011).

\bibitem{ichikawa2014}
T. Ichikawa, Master's Thesis, Tokyo Insutitute of Technology (2014).

\bibitem{JOHANSSON20121760}
J. Johansson, P. Nation, and F. Nori, Comput. Phys. Commun. \textbf{183}, 1760 (2012).

\bibitem{JOHANSSON20131234}
J. Johansson, P. Nation, and F. Nori, Comput. Phys. Commun. \textbf{184}, 1234 (2013).

\end{thebibliography}
\end{document}